\documentclass[12pt]{article}
 \usepackage{epsfig}
 \def\be{\begin{equation}}
 \def\ee{\end{equation}}
 \def\bea{\begin{eqnarray}}
 \def\eea{\end{eqnarray}}
 \usepackage{graphicx}

 \catcode`\@=11
 \def\lsim{\mathrel{\mathpalette\@versim<}}
 \def\gsim{\mathrel{\mathpalette\@versim>}}
 \def\@versim#1#2{\vcenter{\offinterlineskip
 \ialign{$\m@th#1\hfil##\hfil$\crcr#2\crcr\sim\crcr } }}
 \catcode`\@=12

 \catcode`@=12
\begin{document}
 \thispagestyle{empty}
 \begin{flushright}
 UCRHEP-T620\\
 Mar 2022\
 \end{flushright}
 \vspace{0.6in}
 \begin{center}
 {\LARGE \bf Connecting Dark Gauge Symmetry\\
to the Standard Model\\}
 \vspace{1.5in}
 {\bf Ernest Ma\\}
 \vspace{0.1in}
{\sl Department of Physics and Astronomy,\\ 
University of California, Riverside, California 92521, USA\\}
\end{center}
 \vspace{1.2in}

\begin{abstract}\
Dark matter is postulated to be a neutral Dirac fermion, charged under a 
dark $U(1)_D$ gauge symmetry.  Scalar partners of the quarks and leptons 
are also charged under $U(1)_D$.  The dark gauge boson $Z_D$ and the dark 
Higgs boson $h_D$ enable either freeze-out or freeze-in mechanisms to 
account for the correct dark matter relic abundance.  Dark number $D$ is 
connected to baryon number $B$ and lepton number $L$ through 
$D = 3B+L-(2j)_{[mod~2]}$ where $j$ is the intrinsic spin of the particle.
\end{abstract}

\newpage
\baselineskip 24pt
\noindent \underline{\it Introduction}~:~ 
The nature of dark matter~\cite{y17} is unknown, but it is likely to be 
stabilized by an unbroken symmetry, the simplest of which is dark parity, 
under which it is odd and all particles of the Standard Model (SM) are 
even.  In many proposed models involving Majorana neutrinos, it is 
derivable~\cite{m15} from lepton parity, i.e. $\pi_D = \pi_L (-1)^{2j}$, 
where $\pi_L = (-1)^L$ and $j$ is the intrinsic spin of the particle.  
For Dirac neutrinos with the conservation of lepton number $L$, the analog 
connection~\cite{m20} is $D = L-(2j)_{[mod~2]}$.  If baryon number $B$ is 
also considered, it may become $D = 3B+L-(2j)_{[mod~2]}$.  In this paper, 
it is shown how it comes about from a dark gauged $U(1)_D$ symmetry, which 
is broken spontaneously by three units~\cite{mpr13}, instead of the 
customary two. 

To support an anomaly-free $U(1)_D$ gauge symmetry, a neutral Dirac fermion 
$N$ is proposed with dark charge $D=-1$.  It connects to quarks and leptons 
through their scalar counterparts with $D=1$.  The particle content is 
therefore very similar to that of the supersymmetric standard model.  
There are however very important differences.  Instead of two Higgs 
superfields, there is only the one SM Higgs doublet, and there are no 
gauginos.  In contrast, there is a dark gauge boson $Z_D$ and a dark Higgs 
boson $h_D$.  Together with $N$ as dark matter, they participate in 
either freeze-out or freeze-in scenarios.  Whereas $Z_D$ decays immediately 
to $N \bar{N}$ if kinematically allowed, $h_D$ decays mainly through its 
mixing with the SM Higgs boson. 
If $m_{Z_D} < 2 m_N$, then $Z_D$ decays to an SM fermion-antifermion pair, 
through the latter's dark magnetic moment. 
The model and its details are described below.

\noindent \underline{\it Model}~:~ 
The idea of this model and its implementation are both very simple.  
\begin{table}[tbh]
\centering
\begin{tabular}{|c|c|c|c|c|c|c|}
\hline
fermion/scalar & $SU(3)_C$ & $SU(2)_L$ & $U(1)_Y$ & $U(1)_D$ & $B$ & $L$ \\
\hline
$(u,d)_L$ & 3 & 2 & 1/6 & 0 & 1/3 & 0 \\ 
$u_R$ & 3 & 1 & 2/3 & 0 & 1/3 & 0 \\ 
$d_R$ & 3 & 1 & $-1/3$ & 0 & 1/3 & 0 \\ 
\hline
$(\nu,l)_L$ & 1 & $2$ & $-1/2$ & $0$ & $0$ & $1$ \\ 
$\nu_R$ & 1 & $1$ & $0$ & 0 & 0 & $1$ \\ 
$l_R$ & 1 & $1$ & $-1$ & 0 & 0 & $1$ \\ 
\hline
$(\phi^+,\phi^0)$ & 1 & $2$ & $1/2$ & $0$ & $0$ & $0$ \\ 
\hline
$N_{L,R}$ & 1 & $1$ & $0$ & $-1$ & $0$ & $0$  \\ 
\hline
$(\tilde{u}_L, \tilde{d}_L)$ & 3 & 2 & 1/6 & 1 & 1/3 & 0 \\ 
$\tilde{u}_R$ & 3 & 1 & 2/3 & 1 & 1/3 & 0 \\ 
$\tilde{d}_R$ & 3 & 1 & $-1/3$ & 1 & 1/3 & 0 \\ 
\hline
$(\tilde{\nu}_L,\tilde{l}_L)$ & 1 & $2$ & $-1/2$ & $1$ & $0$ & $1$ \\ 
$\tilde{\nu}_R$ & 1 & $1$ & $0$ & 1 & 0 & $1$ \\ 
$\tilde{l}_R$ & 1 & $1$ & $-1$ & 1 & 0 & $1$ \\ 
\hline
$\zeta^0$ & 1 & $1$ & $0$ & $3$ & $0$ & $0$ \\ 
\hline
\end{tabular}
\caption{Fermions and scalars in the dark $U(1)_D$ model.}
\end{table}
All particles of the SM do not transform under $U(1)_D$.  To each quark 
and lepton, a scalar counterpart is added which is charged $+1$ under 
$U(1)_D$, thereby connecting to the neutral Dirac fermion $N$ which is 
charged $-1$.  The neutral scalar $\zeta^0$ breaks $U(1)_D$ by three units. 
It cannot couple to $N_L N_L$ or $N_R N_R$, hence the global symmetry $D$ 
remains, with $N$ having $D=-1$ and the scalar quarks and leptons having 
$D=1$, thereby ensuring the connection $D=3B+L-(2j)_{[mod~2]}$.

\noindent \underline{\it Dark $U(1)_D$ Gauge Symmetry}~:~
Whereas the SM $SU(3)_C \times SU(2)_L \times U(1)_Y$ gauge symmetry is 
broken by the one Higgs doublet $\Phi=(\phi^+,\phi^0)$, the dark 
$U(1)_D$ gauge symmetry is broken by the one Higgs singlet $\zeta^0$. 
The Higgs potential consisting of $\Phi$ and $\zeta$ is thus very simple, i.e. 
\begin{equation}
V = m_1^2 \Phi^\dagger \Phi + m_2^2 \zeta^* \zeta + {1 \over 2} \lambda_1 
(\Phi^\dagger \Phi)^2 + {1 \over 2} (\zeta^* \zeta)^2 + \lambda_3 
(\Phi^\dagger \Phi)(\zeta^* \zeta).
\end{equation}
Let $\langle \phi^0 \rangle = v$ and $\langle \zeta \rangle = u$, then
\begin{equation}
v^2 = {-\lambda_2 m_1^2 + \lambda_3 m_2^2 \over \lambda_1 \lambda_2 
- \lambda_3^2}, ~~~ u^2 = {-\lambda_1 m_2^2 + \lambda_3 m_1^2 \over 
\lambda_1 \lambda_2 - \lambda_3^2}.
\end{equation}
The only physical scalars are the SM $h=\sqrt{2}Re(\phi^0)$ and the new 
dark Higgs boson $h_D=\sqrt{2}Re(\zeta^0)$, and 
\begin{eqnarray}
V &=& \lambda_1 v^2 h^2 + \lambda_2 u^2 h_D^2 + 2 \lambda_3 v u h h_D + 
{1 \over \sqrt{2}} \lambda_1 v h^3 + {1 \over \sqrt{2}} \lambda_2 u h_D^3 
\nonumber \\ &+& {1 \over \sqrt{2}} \lambda_3 v h h_D^2 + {1 \over \sqrt{2}} 
\lambda_3 u h_D h^2 + {1 \over 8} \lambda_1 h^4 + {1 \over 8} \lambda_2 h_D^4 
+ {1 \over 4} \lambda_3 h^2 h_D^2.
\end{eqnarray}
Hence $h_D$ mixes with $h$ according to
\begin{equation}
M^2_{h h_D} = \pmatrix{2 \lambda_1 v^2 & 2 \lambda_3 v u \cr 2 \lambda_3 v u 
& 2 \lambda_2 u^2},
\end{equation}
and decays to SM fermions through the $h$ Yukawa couplings.  Note that 
$\zeta^0$ has $D=3$, hence $h_D$ does not couple to $N_L N_L$ or 
$N_R N_R$.
 
As for $Z_D$, the gauge invariant term  
$|(\partial^\mu - 3ig_D Z_D^\mu)\zeta|^2$ yields $m_{Z_D}=3\sqrt{2}g_Du$, 
together with the interaction
\begin{equation}
{\cal L}_{int} = 9\sqrt{2}g_D^2u h_D Z_D^\mu {Z_D}_\mu + {9 \over 2} g_D^2 
h_D^2 Z_D^\mu {Z_D}_\mu.
\end{equation}
This means that $h_D$ may decay into $Z_D Z_D$ if kinematically allowed.
It also couples to $N \bar{N}$ indirectly through $Z_D$ in one loop.

Since $N$ is charged under $U(1)_D$, as are the scalar quarks and leptons, 
they are decay products of $Z_D$ if kinematically allowed.  If not, then 
$Z_D$ may decay into an SM fermion-antifermion pair.  The $\gamma^\mu$ 
\begin{figure}[htb]
 \vspace*{-5cm}
 \hspace*{-5cm}
 \includegraphics[scale=1.0]{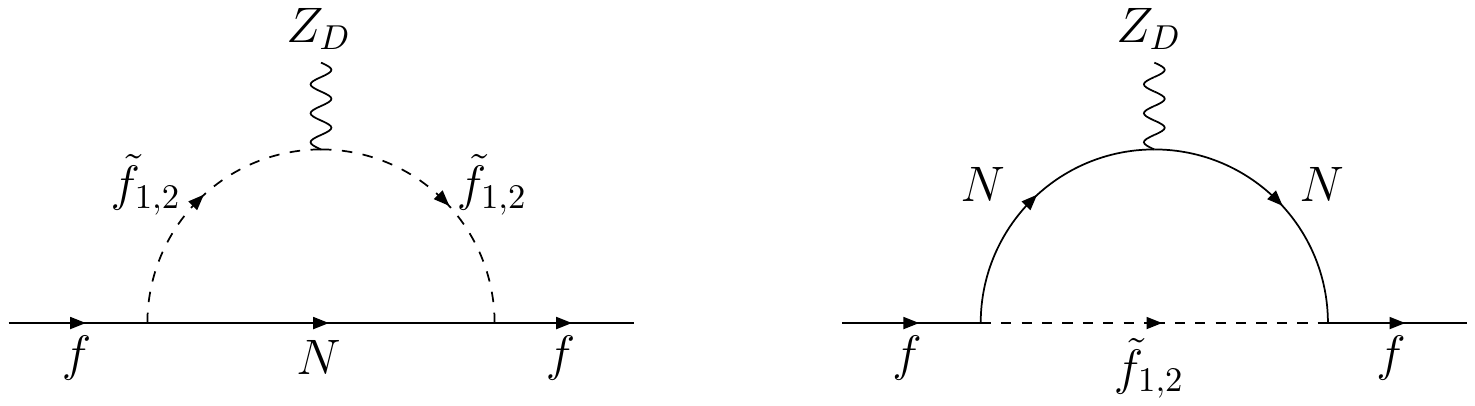}
 \vspace*{-20.5cm}
 \caption{Dark magnetic moment of SM fermion.}
 \end{figure}
coupling is zero from dark gauge invariance.  The $\sigma^{\mu \nu}$ 
coupling is obtained as shown in Fig.~1, where $\tilde{f}_{1,2}$ are the mass 
eigenstates from $\tilde{f}_L-\tilde{f}_R$ mixing.
Let $\tilde{f}_L = \cos \theta \tilde{f}_1 - \sin \theta \tilde{f}_2$ and  
$\tilde{f}_R = \sin \theta \tilde{f}_1 + \cos \theta \tilde{f}_2$,
with $m^2_{1,2} >> m_N^2 >> m_f^2$, then the dark magnetic moment of $f$ 
is~\cite{qs14}
\begin{equation}
a_f = { g_f^2 \sin \theta \cos \theta \over 4 \pi^2} m_f m_N \left[ 
{1 \over m_1^2} \left( \ln {m_1^2 \over m_N^2} - {7 \over 4} \right) - 
{1 \over m_2^2} \left( \ln {m_2^2 \over m_N^2} - {7 \over 4} \right) \right], 
\end{equation}
where $g_f^L = g_f^R$ has been assumed.  The correspinding interaction is 
\begin{equation}
{\cal L}_{int} = {i g_D a_f \over 2m_f} Z_D^\mu q^\nu 
\bar{f} \sigma_{\mu \nu} f.
\end{equation}

\noindent \underline{\it Scotogenic Fermion Masses}~:~
According to Table~1, quarks and leptons obtain masses as in the SM. 
However, it is a simple matter to allow some light fermions, such as the 
Dirac neutrinos, to acquire radiative masses through dark matter~\cite{m06}, 
by the implementation of a softly broken $Z_2$ symmetry.  For example, 
let $\nu_R$ and $\tilde{\nu}_R$ be odd under this $Z_2$, then the 
dimension-four Yukawa coupling $\bar{\nu}_R (\nu_L \phi^0 - l_L \phi^+)$ 
is forbidden, but $\bar{\nu}_R N_L \tilde{\nu}_R$ is allowed.  Hence 
\begin{figure}[htb]
 \vspace*{-5cm}
 \hspace*{-3cm}
 \includegraphics[scale=1.0]{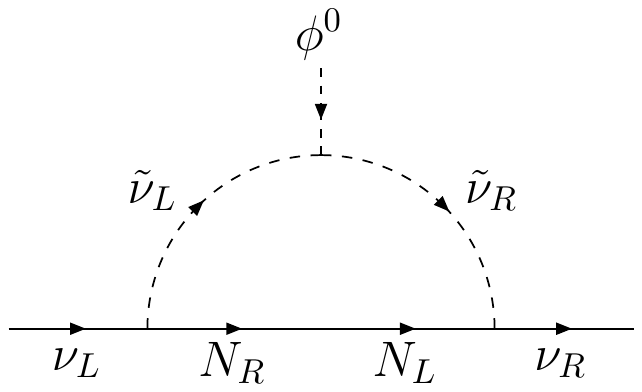}
 \vspace*{-21.5cm}
 \caption{Scotogenic Dirac neutrino mass.}
 \end{figure}
the one-loop generation of Dirac neutrino mass is possible as shown in 
Fig.~2, where the soft breaking of $Z_2$ occurs with the scalar trilinear 
$\tilde{\nu}_R^* (\phi^0 \tilde{\nu}_L - \phi^+ \tilde{l}_L)$ term. 
Analogous constructions are possible for the electron and muon masses 
if desired.  Note that $U(1)_D$ is also applicable to the dark sector 
of the recently proposed~\cite{m22} scotogenic $A_5 \to A_4$ model of 
Dirac neutrino masses.

\noindent \underline{\it Freeze-Out Scenario}~:~
The neutral dark fermion $N$ is a candidate for the dark matter of the 
Universe from thermal freeze-out.  Assuming that $m_N > m_{Z_D}$, the 
annihilation of $N \bar{N} \to Z_D Z_D$ is shown in Fig.~3.
\begin{figure}[htb]
 \vspace*{-5cm}
 \hspace*{-3cm}
 \includegraphics[scale=1.0]{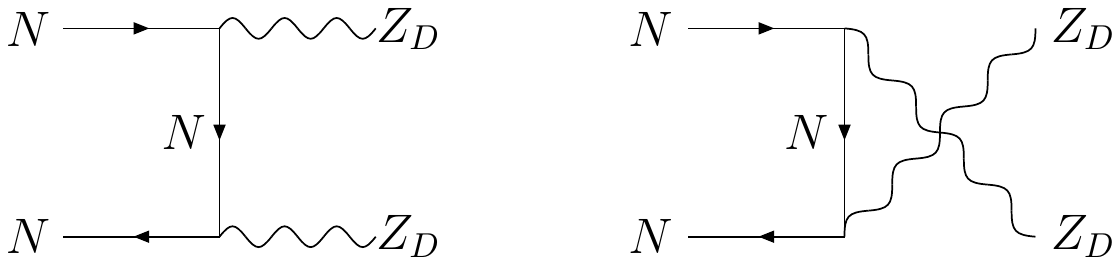}
 \vspace*{-21.5cm}
 \caption{Annihilation of $N \bar{N} \to Z_D Z_D$.}
 \end{figure}
The cross section $\times$ relative velocity is given by~\cite{m21}
\begin{equation}
\sigma v_{rel} = {g_D^4 \over 16 \pi m_N^2} \left( 1 - {m_{Z_D}^2 \over m_N^2} 
\right)^{3 \over 2} \left( 1 - {m_{Z_D}^2 \over 2m_N^2} 
\right)^{- 2}. 
\end{equation}
Setting this value to the canonical $6 \times 10^{-26}~{\rm cm}^3/{\rm s}$ 
for a Dirac fermion, and assuming $m_N = 1$ TeV and $m_{Z_D} = 800$ GeV, it 
is satisfied for $g_D=0.86$.  Once produced, $Z_D$ thermalizes with the SM 
particles through $h_D$ and its interactions listed in Eqs.~(3) and (5). 

As for direct detection, $N$ interacts with quarks through their scalar 
counterparts.  For simplicity, let $g_u^L=g_u^R=g_d^L=g_d^R=g_0$ and 
$m_{\tilde{u}_L}=m_{\tilde{u}_R}=m_{\tilde{d}_L}=m_{\tilde{d}_R}=\tilde{m}_0$, 
then the elastic scattering of $N$ off a Xenon nucleus per nucleon is given by 
\begin{equation}
\sigma_0 = {g_0^4 \mu^2 \over 64 \pi (\tilde{m}_0^2 - m_N^2)^2},
\end{equation}
where $\mu=m_N m_{Xe}/(m_N+m_{Xe})$ is the reduced mass of $N$. For $m_N=1$ 
TeV, and using $m_{Xe}=122.3$ GeV with 
$\sigma_0 < 10^{-45}$~cm$^2$~\cite{xenon18}, the lower limit on 
$\tilde{m}_0/g_0$ is about 70 TeV.  Note that this severe constraint 
rules out the possibility that the relic abundance of $N$ comes from 
this interaction.  Without the dark gauge $U(1)_D$ symmetry, $N$ would 
not be a dark matter candidate in this case.

\noindent \underline{\it Freeze-In Scenario}~:~
An alternative for $N$ to be dark matter is the 
freeze-in mechanism~\cite{hjmw10}.  Here $N$ is assumed to be light,  
and its only production is through Higgs decay~\cite{m19} as shown in Fig.~4.  
\begin{figure}[htb]
 \vspace*{-5cm}
 \hspace*{-5cm}
 \includegraphics[scale=1.0]{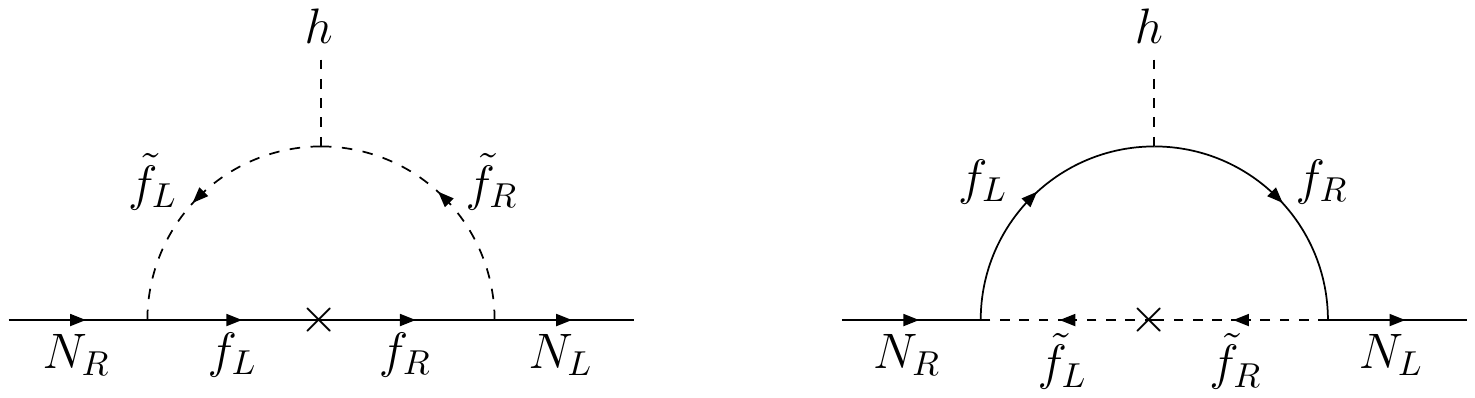}
 \vspace*{-20.5cm}
 \caption{Higgs decay to $N \bar{N}$.}
 \end{figure}
This scenario works if the reheat temperature of the Universe is 1 to 10 TeV, 
and the thermal production of $N$ is very much suppressed, so that its 
relic abundance comes only from Higgs decay until the latter goes out 
of thermal equilibrium with the other SM particles.
The effective $h$ coupling to $\bar{N} N$ is 
\begin{equation}
g_h = {g_f^2 \mu_f m_f \over 8\pi^2} F(\tilde{m}_f^2,m_f^2),
\end{equation}
where $F(x,y)= (x-y)^{-1}-y\ln(x/y)(x-y)^{-2}$, and $\mu_f$ is the 
$h \tilde{f}_L^* \tilde{f}_R$ coupling.  The decay rate of $h \to N \bar{N}$ 
is then
\begin{equation}
\Gamma_h = {g_h^2 m_h \over 8 \pi} \sqrt{1-4r^2} (1-2r^2),
\end{equation}
where $r=m_N/m_h$.  The correct relic abundance is possible if $g_h$ is 
very small.  Hence $N$ could be FIMP (Feebly Interacting Massive Particle) 
dark matter~\cite{hjmw10}, and for $r<<1$, the right number density is 
obtained for~\cite{ac13}
\begin{equation}
g_h \sim 10^{-12}r^{-1/2}.
\end{equation}
Assuming $m_N=2$ GeV and $\tilde{m}_f/g_f=10^5$ GeV~\cite{mr21}, this would 
require $\mu_f m_f \sim 6$ GeV$^2$.  In this scenario, both $Z_D$ and $h_D$ 
should also be much heavier than $h$.

\noindent \underline{\it Possible Light $Z_D$ and $h_D$}~:~
In the freeze-out scenario, it is possible to have light $Z_D$ and $h_D$, 
such that $h$ decays to $h_D h_D$, then $h_D$ decays to $Z_D Z_D$, and 
$Z_D$ decays to an SM fermion-antifermion pair.  This would result in 
multi-lepton final states of the SM Higgs boson decay, as pointed 
out previously~\cite{cmy14}.  The respective decay rates are
\begin{eqnarray}
\Gamma(h \to h_D h_D) &=& {\lambda_3^2 v^2 \over 16 \pi m_h} 
\sqrt{1-{4 m_{h_D}^2 \over m_h^2}}, \\
\Gamma(h_D \to Z_D Z_D) &=& {9 g_D^2 \over 16 \pi m_{h_D}} 
(m^2_{h_D} + 2 m^2_{Z_D})\sqrt{1-{4 m_{Z_D}^2 \over m_{h_D}^2}}, \\
\Gamma(Z_D \to f \bar{f}) &=& {g_D^2 a_f^2 \over 96 \pi m_{Z_D} m_f^2} 
(m^2_{Z_D} + 2 m^2_f)(m^2_{Z_D} + 8 m^2_f) 
\sqrt{ 1 - {4m_f^2 \over m^2_{Z_D}}}.
\end{eqnarray}

\noindent \underline{\it Concluding Remarks}~:~
In the SM, fermions dominate in the form of quarks and leptons.  Bosons 
appear as the necessary vector gauge particles of the SM 
$SU(3)_C \times SU(2)_L \times U(1)_Y$ gauge symmetry, and only one physical 
scalar Higgs boson remains.  It is postulated here that in the dark sector, 
the opposite occurs, with one neutral Dirac fermion $N$ and many scalars, 
i.e. the scalar partners of the SM quarks and leptons.  They are charged 
under a dark $U(1)_D$ gauge symmetry, which is broken spontaneously 
through a neutral scalar by 3 units, resulting in one dark gauge boson 
$Z_D$ and one dark Higgs boson $h_D$.  Consequently, a residual global 
dark number is identified as $D = 3B+L-(2j)_{[mod~2]}$, where $j$ is the 
intrinsic spin of the particle.  The combined SM and dark sectors are also 
very suitable for obtaining scotogenic Dirac neutrino masses with the 
imposition of a softly broken $Z_2$ symmetry.

The stable dark matter is $N$.  It may be thermally produced and 
annihilates to $Z_D$ in the freeze-out mechansim, or through the rare 
decay of the SM Higgs boson $h$ in the freeze-in mechanism.  In the 
former case, $h$ may decay to $h_D h_D$, then $h_D$ to $Z_D Z_D$, then 
$Z_D$ to an SM fermion-antifermion pair, resulting in multilepton final 
states.

\noindent \underline{\it Acknowledgement}~:~
This work was supported in part by the U.~S.~Department of Energy Grant 
No. DE-SC0008541.  

\bibliographystyle{unsrt}

\end{document}